\documentclass[12pt,a4paper]{article}

\usepackage{latexsym}

\title{Comparing metrics at large:\\ harmonic vs quo-harmonic coordinates}

\author{J.\ M.\ Aguirregabiria%
\footnote{\ Tel. +34 946012593, Fax: +34 944648500, e-mail: \texttt{wtpagagj@lg.ehu.es}}
${}^1$, Ll.\ Bel${}^1$, J.\ Mart\'{\i}n${}^2$,\\ 
A.\ Molina${}^3$ and E.\ Ruiz${}^2$
\\${}^1$\emph{Fisika Teorikoa, Euskal Herriko Unibertsitatea},
\\\emph{P.K. 644, 48080 Bilbo, Spain}
\\${}^2$\emph{Area de F\'\i sica Te\'orica. Edificio Triling\"ue},
\\\emph{Universidad de Salamanca. 37008 Salamanca, Spain}
\\${}^3$\emph{Dep.\ de F\'\i sica Fonamental, Universitat de Barcelona},
\\\emph{Diagonal 647, Barcelona 08028 i Societat Catalana de F\'\i sica}
}

\begin{document}

\maketitle

\date{}

\begin{abstract}

To compare two space-times on large domains, and in particular the
global structure of their manifolds, requires using identical frames of
reference and associated coordinate conditions. In this paper we use
and compare two classes of time-like congruences and corresponding
adapted coordinates: the harmonic and quo-harmonic classes. Besides the
intrinsic  definition and some of their intrinsic properties and
differences we consider with some detail their differences at the level
of the linearized approximation of the field equations. The hard part of
this paper is an explicit and general determination of the harmonic and
quo-harmonic coordinates adapted to the stationary character of three
well-know metrics, Schwarzschild's, Curzon's and Kerr's, to  order five
of their asymptotic expansions. It also contains some relevant remarks
on such problems as defining the multipoles of vacuum solutions or
matching interior and exterior solutions.

\end{abstract}

\newpage

\section{Introduction}\label{sec:intro} 
Let us consider  three well-known vacuum solutions of Einstein's
equations: 

1.~The Schwarzschild solution written in Droste-Hilbert coordinates:
\begin{equation}
\label{eq:1.1}
ds_S^2=-\left(1-\frac{2M}{\bar{r}}\right)\,dt^2+
\frac1{1-2M/\bar{r}}\,d\bar{r}^2+
\bar{r}^2(d\bar\theta^2+\sin^2\bar\theta\, d\bar\varphi^2),
\end{equation}
\indent with
\begin{equation}
\label{eq:1.6}
\bar r > 2M.
\end{equation}

2.~The Curzon solution written in spherical Weyl-related coordinates:
\begin{eqnarray}
\label{eq:1.2}
ds_C^2=&-&e^{-2M/\tilde{r}}\,dt^2\nonumber \\
&+&e^{2M/\tilde{r}}
\left[
\exp\left(-\frac{M^2\sin^2\tilde\theta}{\tilde{r}^2}\right)
(d\tilde{r}^2+\tilde{r}^2\,d\tilde{\theta}^2)+
\tilde{r}^2\sin^2\tilde\theta\, d\tilde\varphi^2\right],
\end{eqnarray} 
\indent with
\begin{equation}
\label{eq:1.7}
\tilde r>0.
\end{equation}

3.~The Kerr solution written in Boyer-Lindquist coordinates,
\begin{eqnarray}
\label{eq:1.3}
ds_K^2=&-&\left(1-\frac{2M\check r}{\sigma^2}\right)\,dt^2
+\frac{4M\check{r}}{\sigma^2}a\sin\check\theta^2\,dt\,d\check\varphi \nonumber \\
&+&\frac{\sigma^2}{\Delta}\,d\check{r}^2+\sigma^2\,d\check\theta^2+
\left(\check{r}^2+a^2+\frac{2M\check{r}}{\sigma^2}a^2
\sin\check\theta^2\right)\sin\check\theta^2\,d\check\varphi^2,
\end{eqnarray}
\indent with
\begin{equation}
\label{eq:1.4}
\sigma^2=\check{r}^2+a^2\cos\check\theta^2 > 2M\check r, \qquad
\Delta=\check{r}^2+a^2-2M\check{r} >0.
\end{equation}

We have used different notations to distinguish the radial and 
angular coordinates to emphasize that there is no connection {\it a priori} that
makes sense across the three solutions.

These three space-time metrics share some intrinsic properties, 
i.e.\ which are independent of the system of coordinates being used to
describe them. Namely: the three solutions possess a main global
time-like Killing vector~$\xi^\alpha$ in the domains specified in
Eqs.~(\ref{eq:1.6}), (\ref{eq:1.7}) and~(\ref{eq:1.4}). 
This vector is actually a
generator of a group of isometries which includes as sub-group the
group of rotations around an axis, and the specular symmetry across a
plane. The intrinsic differences are also well known. Namely: the
Schwarzschild and Curzon solutions are static, i.e.\ $\xi^\alpha$ is
integrable, while the Kerr solution is only stationary. The Schwarzschild
solution is spherically symmetric while the Curzon and the Kerr
solutions are only axially symmetric.

The three systems of coordinates used in Eqs.~(\ref{eq:1.1}), (\ref{eq:1.2}) 
and~(\ref{eq:1.3})
share also some common properties:

i) The space coordinates are adapted to the main time-like
Killing vector $\xi^\alpha$ as well as to the space-like Killing vector
$\zeta^\alpha$ corresponding to the axial symmetry. This is reflected
by the fact that the gravitational potentials are independent of $t$
and $\varphi$. They are also adapted to the specular symmetry of 
the solutions.   

ii) They clearly show their Euclidean
behavior at spatial infinity in the frame of reference corresponding
to the congruence defined by $\xi^\alpha$.

These common properties by no means make these coordinates unique,
and therefore it would be even heuristically unjustified to refer to
these three metrics as being written using a ``common" system of
coordinates. This paper deals with the problem of further restricting
the systems of coordinates up to the point of making such an
assertion acceptable.

We shall see that the solution to this problem sheds new light on the
problem 
of matching them to interior solutions. On
the other hand it is obvious that being able to use a ``common"
system of coordinates is the only available means to compare
different solutions on large domains of their manifolds.

Coordinate specifications can be made on different criteria. The Droste-Hilbert
coordinates in Eq.~(\ref{eq:1.1}) are completely determined by the
requirement
\begin{equation}
\label{eq:1.8}
ds^2_S\mid_\Sigma={\bar r}^2(d{\bar\theta}^2+\sin^2\bar\theta
d\bar\varphi^2),
\end{equation}
$\Sigma$ being any of the 2-surfaces $t=\mbox{const.}$, $\bar r=\mbox{const.}$ It
can be used only for spherically symmetric space-times. The Weyl-like
coordinates in Eq.~(\ref{eq:1.2}) are specially tailored for static axially
symmetric metrics. The Schwarzschild solution could be written using such
coordinates but we all know how misleading this
would be from a physical point of view. 

The Boyer-Lindquist coordinates are akin to Droste-Hilbert coordinates in
the sense that when the parameter $a$ in the Kerr solution is
made zero then Eq.~(\ref{eq:1.3}) becomes Eq.~(\ref{eq:1.1}).

It is clear that the connections between the three systems of
coordinates we have considered are as yet too loose and partial to be 
of any interest. Above all because they do not shed any new light on
any other problem. 

We take in this paper the point of view according to which 
to determine the physical content of the space-time metric being considered
as well as to be able to compare
two of them on large domains it is necessary to use systems of
coordinates whose definition make sense independently of the metric
to which the definition is applied.

In Section~\ref{sec:harmvsquo} we consider the definition and some properties 
of harmonic and quo-harmonic congruences and the intrinsically related
harmonic and quo-harmonic coordinates.
The linear approximation discussed in Section~\ref{sec:linear} provides a simplified
framework to illustrate some interesting properties 
of those congruences.

Section~\ref{sec:Schw} deals with the problem of writing the Schwarzschild solution, 
up to terms of order $M^5$ included, both in harmonic and quo-harmonic
coordinates adapted to the intrinsic symmetry structure described above.

In Section~\ref{sec:Curzon} we consider the same problem for the Curzon solution. Our
results show in particular that Schwarzschild and Curzon
solutions coincide when terms of order $M^3$ and higher can be
neglected. This is the case when $M$ is the mass of the Sun and the
physical system being considered is the solar system. 
  
The same problem for the Kerr solution is analysed in Section~\ref{sec:Kerr}. Our
results show in particular that when considering asymptotic
developments in $1/r$ of this metric using quo-harmonic coordinates it is 
necessary to include some logarithmic terms behaving as $r^{-5}\ln r$.


\section{Harmonic and quo-harmonic coordinates}\label{sec:harmvsquo}

Whenever the metric 
\begin{equation}
ds^2=g_{\alpha\beta}\,dx^\alpha dx^\beta
\end{equation}
of a space-time model is given using an explicit
system of coordinates $x^\alpha$, two associated geometric structures are also implicitly
given: i) the congruence $\cal C$ of world-lines defined by the parametric
equations
\begin{equation}
\label{eq:2.1}
x^i=\mbox{constant}, \quad i=1,2,3,
\end{equation}  
which we shall always assume to be time-like in the domain $\cal D$ of
interest, and ii) the foliation $\mathcal{F}$ defined by the equation
\begin{equation}
\label{eq:2.2}
x^0=\mbox{constant},
\end{equation}
which we shall always assume to be space-like on $\cal D$.

Two questions can be asked which are relevant in the theory of frames
of reference or more generally whenever we are interested in the
suitability of the coordinate system to handle a particular problem:
i) does the congruence $\mathcal{C}$ belong to any particular type, like
being for instance a Killing or Born or any other type intrinsically
defined?  and ii) can the foliation $\mathcal{F}$ be intrinsically
characterized independently of $\mathcal{C}$, or does it have any
particular intrinsic properties connected with it?

Systems of coordinates can be divided in two classes: those that are
specific to particular models or classes of models and are useful
because they take advantage of particular properties of them, and
those that make sense whatever model is being considered. Isotropic, Droste-Hilbert
or Schwarzschild coordinates for space-times with spherical
symmetry, or Weyl coordinates for static axially symmetric models
belong to the first class.

Gauss coordinates belong to the second class, but to our
knowledge the single problem where they have played an important role
in the development of the theory of General Relativity is in the
problem of matching exterior to interior solutions of Einstein's
equations across space-like hypersurfaces. More on
that  on Section~\ref{sec:matching}.

\subsection{Harmonic congruences and coordinates}\label{sec:harmonic}

A widely used system of coordinates $x^\alpha$ belonging to this second class is
that of harmonic coordinates, which is characterized by these two
groups of equations:
\begin{equation}
\label{eq:2.3}
\Box x^i=0
\end{equation}
and
\begin{equation}
\label{eq:2.4}
\Box x^0=0,
\end{equation}     
where $\Box$ is the intrinsic d'Alembertian of the space-time
metric. The splitting above in two groups of the four equations $\Box
x^\alpha=0$ will be understood in a moment.

Conversely, given a time-like congruence defined by a unit tangent
vector field $u^\alpha$ it may be asked what type of adapted coordinates can
be  most appropriate to use. For instance, we know that if $u^\alpha$
is collinear with a Killing vector field then it is always possible to use a
system of harmonic coordinates such that the congruence with parametric equations
(\ref{eq:2.1}) is the same as that defined by $u^\alpha$. But in general
this is not the case, because if we consider the harmonic
coordinates $x^i$ as functions of generic coordinates, 
$x^i=f^i(y^\alpha)$, then the functions $f^i$ must satisfy the system of equations 
\begin{equation}
\label{eq:2.5}
\Box f^i=0, \quad u^\alpha\partial_\alpha f^i=0, \quad 
\hbox{rank}(\partial_\alpha f^i)=3,
\end{equation}
and this requires integrability conditions on the vector field $u^\alpha$
that in general are not satisfied. A particularly interesting case
in which they do not hold is
that of irrotational Born (or rigid) congruences when they are not
Killing~\cite{Tolo}. 
As a matter of fact, Eqs.~(\ref{eq:2.5}) can be considered as the
definition of a new intrinsic type of congruences: that of harmonic
congruences, i.e.\ those admitting a set of adapted
harmonic coordinates of space.

Notice that the time coordinate can always be required to satisfy 
Eq.~(\ref{eq:2.4})
and therefore this condition is, without any
supplementary conditions, unrelated to any particular congruence:
it is enough to find a solution of
\begin{equation}
\Box f^0=0, \qquad u^\alpha\partial_\alpha f^0\ne0.
\end{equation}
It is
for this reason that we preferred to refer to harmonic coordinates
splitting the conditions on the space and time coordinates.

\subsection{Quo-harmonic congruences and coordinates}\label{sec:quoharmonic}

As we have already mentioned these considerations are particularly
relevant in any theory of frames of reference that requires to extend
this concept beyond the Born congruences which are, generically,
notoriously exceptional. As a contribution towards an appropriate 
generalization it has been introduced~\cite{Salas,Llosa1,Llosa2}
 a new type of congruences which
are defined much in the same way as we defined harmonic congruences
by Eqs.~(\ref{eq:2.5}), except for a slight modification of the first group
of conditions, which becomes
\begin{equation}
\label{eq:2.6}
\Box f^i+\Lambda^\alpha\partial_\alpha f^i=0,
\end{equation}     
where
\begin{equation}
\label{eq:2.7}
\Lambda^\alpha=-u^\rho\nabla_\rho u^\alpha
\end{equation} 
is, up to the sign, the intrinsic curvature of the congruence
$u^\alpha$.
We shall also consider later a quo-harmonic time coordinate
defined by
\begin{equation}
\label{eq:2.60}
\Box f^0+\Lambda^\alpha\partial_\alpha f^0=0,\qquad
u^\alpha\partial_\alpha f^0\ne0.
\end{equation} 

This class of quo-harmonic congruences contains, in contradistinction
to the class of harmonic congruences, the whole class of Born
congruences. And it provides sufficient generality to be an essential
ingredient to define the concept of rigidity without which
no useful meaning can be given to the concept of frame of reference.  
This is why we shall spend some time to justify the consideration of
quo-harmonic congruences and quo-harmonic coordinates; two concepts
that are related but that by no means are identical.

\subsection{Harmonic and quo-harmonic coordinate classes}

The next section is dedicated to compare Einstein's linearized equations
using harmonic or quo-harmonic coordinates. In the remaining sections
instead we shall restrict ourselves to the consideration of the three
stationary metrics mentioned in the Introduction. 
These metrics are stationary and have been written
in a system of coordinates adapted to a main Killing vector.
All Killing congruences are both harmonic and quo-harmonic. This can 
be seen as follows. In a system of coordinates adapted to a Killing vector
field all the potential $g_{\alpha\beta}$ are
independent of time and Eqs.~(\ref{eq:2.6}) become
\begin{equation}\label{eq:deltafk}
\hat\Delta\hat f^k=\frac{1}{\sqrt{\hat g}}\partial_i\left(\sqrt{\hat g}\,\hat g^{ij}
\partial_j \hat f^k\right)=0,\qquad\hat g\equiv\det\left(\hat g_{ij}\right),
\end{equation}
where it has been used the decomposition
\begin{equation}
ds^2=-\left[\xi\left(-\,dt+\varphi_i\,dx^i\right)\right]^2+\hat g_{ij}\,dx^idx^j
\end{equation}
with
\begin{equation}
\xi\equiv\sqrt{-g_{00}},\qquad\varphi_i\equiv\xi^{-2}g_{0i},\qquad
\hat g_{ij}\equiv g_{ij}+\xi^2\varphi_i\varphi_j.
\end{equation}
Since Eqs.~(\ref{eq:deltafk}) always have~\cite{bers,deturk} three solutions
independent of $t$ it follows that the Killing
congruences are quo-harmonic. On the other hand Eqs.~(\ref{eq:2.5})
become 
\begin{equation}\label{eq:deltafbar}
\bar\Delta\bar f^k=\frac{1}{\sqrt{\bar g}}\partial_i\left(\sqrt{\bar g}\,\bar g^{ij}
\partial_j \bar f^k\right)=0,
\qquad \bar g_{ij}\equiv\xi^2\hat g_{ij},
\qquad\bar g\equiv\det\left(\bar g_{ij}\right),
\end{equation}
and these equation also have three independent solutions independent of $t$. 
Therefore Killing congruences are also harmonic.

Notice however that if $\Lambda_i\ne0$ harmonic coordinates
are different from the quo-harmonic ones even for a congruence
that is both harmonic and quo-harmonic 
---a case that includes as we have seen the Killing congruences---
in which case they  reveal different aspects of the three models
being considered. 

The harmonic coordinates are derived from
those used in Eqs.~(\ref{eq:1.1}), (\ref{eq:1.2}) and~(\ref{eq:1.3})
by a pure space transformation
\begin{equation} 
\label{eq:1.9} 
x^i=x^i(y^j,\lambda,\mu),
\end{equation} 
where $y^j$ is any of the systems of coordinates
used in~(\ref{eq:1.1}), (\ref{eq:1.2}) 
and~(\ref{eq:1.3}), $\lambda$~is the set of parameters on which depends the
metric ($M$, but also $a$ for the Kerr metric) and $\mu$ is any set of
constants coming from integrating the coordinate conditions being
demanded. For $x^i$ to be harmonic and Cartesian at space infinity
the gravitational potentials $g_{\alpha\beta}(t,x^i)$ have to be
solutions of the equations
\begin{equation}
\label{eq:1.10}
\Gamma^k\equiv g^{\alpha\beta}\Gamma^k_{\alpha\beta}=0,
\end{equation} 
which are Eqs.~(\ref{eq:2.3}) when harmonic coordinates are used.
Here $\Gamma^k_{\alpha\beta}$ are the Christoffel symbols.
Of course any other system of coordinates~$z^i$ 
derived from $x^i$ by 
\begin{equation} 
\label{eq:1.12} 
z^i= f^i(x^j) 
\end{equation}
with functions $f^i$ independent of the
parameters $\lambda$ and $\mu$ can be said to belong to the harmonic
class and be used to fulfill the requirement of universality mentioned
before. In particular, we shall systematically use not harmonic Cartesian-like
coordinates but rather polar coordinates that belong to the same
harmonic class and are related to them by the
familiar formulas:
\begin{equation}
\label{eq:1.13}
x^1=r\sin\theta\cos\varphi, \quad x^2=r\sin\theta\sin\varphi, \quad x^3=r\cos\theta.
\end{equation}

We will also use quo-harmonic coordinates, which are
derived from the original ones by space-like
transformations of the type~(\ref{eq:1.9}) but requiring instead of~(\ref{eq:1.10})
Eqs.~(\ref{eq:2.6}), which  become
\begin{equation}
\label{eq:1.14}
\hat\Gamma^k\equiv\Gamma^k+g^{kj}\Lambda_j=0
\end{equation}
when quo-harmonic coordinates are used.
For stationary metrics these equations reduce to
\begin{equation}
\label{eq:1.14bis}
\Gamma^k-g^{kj}\partial_j\ln\xi=0.
\end{equation}

\section{The linear approximation}\label{sec:linear}

We assume in this section that the metric admits a congruence
$u^\alpha$ and a system of adapted coordinates $x^\alpha$ such that
it can be written as
\begin{equation}
\label{eq:3.1}
g_{\alpha\beta}=\eta_{\alpha\beta}+h_{\alpha\beta},
\end{equation}
where $\eta_{\alpha\beta}$ are the Galilean coefficients of the
Minkowski metric, and $h_{\alpha\beta}$ are small quantities whose
products can be neglected, as well as the products of derivatives of any
order, in the domain $\mathcal{D}$ of interest.

Adapted coordinate transformations leaving form-invariant (\ref{eq:3.1})
are up to a Lorentz transformation
\begin{equation}
\label{eq:3.2}
{x^\prime}^0=x^0+\zeta^0(x^\alpha), \quad 
{x^\prime}^i=x^i+\zeta^i(x^j),
\end{equation}   
where the $\zeta^\alpha$ are small quantities, and the $\zeta^i$ do not
depend on time. 

Under such an adapted coordinate transformation the quantities
$h_{\alpha\beta}$ transform as follows:
\begin{eqnarray}
\label{eq:3.3}
{h^\prime}_{00}&=&h_{00}+2\partial_0\zeta_0, \\
\label{eq:3.3a}
{h^\prime}_{0i}&=&h_{0i}+\partial_i\zeta_0, \\
{h^\prime}_{ij}&=&h_{ij}+\partial_i\zeta_j+\partial_j\zeta_i.
\end{eqnarray}

The differential invariants under these transformations are
\begin{eqnarray}\label{eq:inv1}
\Lambda_i&=&\frac12\partial_ih_{00}-\partial_0h_{0i}, \\
\Omega_{ij}&=&\partial_ih_{0j}-\partial_jh_{0i}, \\
\Sigma_{ij}&=&\partial_0h_{ij}, \\
\hat R_{ijkl}&=&-\frac12(\partial_{ik}h_{jl}+\partial_{jl}h_{ik}
-\partial_{il}h_{jk}-\partial_{jk}h_{il}).\label{eq:inv4}
\end{eqnarray}
These expressions give in this approximation the values of intrinsically
well defined objects associated to the time-like congruence~\cite{Llosa1,Llosa2}:
the sign-reversed acceleration~(\ref{eq:2.7}), the vorticity field, the deformation rate
and the Zel'manov-Cattaneo tensor~\cite{zelmanov,cattaneo}.

\subsection{Harmonic congruences}\label{sec:linharm}

Let us consider the quantities defined in~(\ref{eq:1.10}), which in this approximation
reduce to
\begin{equation}
\label{eq:3.4}
\Gamma_i=
-\partial_0h_{0i}+\partial_jh^j_i-\frac12\partial_i(-h_{00}+\hat h), 
\end{equation}
where
\begin{equation}
\label{eq:3.5}
\hat h= \delta^{ij}h_{ij}.
\end{equation} 
Under an adapted coordinate transformation those quantities and their
de\-rivatives with respect to time transform as follows:
\begin{equation}
\label{eq:3.6}
\Gamma_{i^\prime}=\Gamma_i+\triangle\zeta_i, \quad 
\partial_0\Gamma_{i^\prime}=\partial_0\Gamma_i.
\end{equation} 
Since at this approximation a congruence is harmonic {\it iff} a system
of coordinates exists such that $\Gamma_{i^\prime}=0$, it follows that
\begin{equation}
\label{eq:3.7}
\partial_0\Gamma_j=\partial_i\Sigma^i_j-\frac12\partial_j\Sigma
+\partial_0\Lambda_j=0,\qquad \Sigma\equiv\Sigma^i_i
\end{equation}
characterizes the harmonic congruences to this approximation. This is in fact an invariant condition 
that guarantees the existence
of a solution of the equation
\begin{equation}
\label{eq:3.8}
\triangle\zeta_i=-\Gamma_i
\end{equation}
for $\zeta_i$ not depending of $x^0$.

If the harmonic congruence is a Born congruence then $\Sigma_{ij}=0$ and it
follows from (\ref{eq:3.7}) that $\Lambda_i$ must be independent of time.
If moreover we assume that $\Omega_{ij}$ is also independent of time
then it is easy to show that the congruence is in fact a Killing
congruence, i.e.\ a system of adapted coordinates exist such that
$\partial_0g_{\alpha\beta}=0$.
                               
Let us now consider the linearized Einstein's field equations,
\begin{equation}
\label{eq:3.13}
R_{\alpha\beta}=U_{\alpha\beta}, \quad 
U_{\alpha\beta}\equiv
T_{\alpha\beta}
-\frac12(\eta^{\rho\sigma}T_{\rho\sigma})\eta_{\alpha\beta},
\end{equation}
where $R_{\alpha\beta}$ is the linearized Ricci tensor:
\begin{equation}
\label{eq:3.14}
R_{\alpha\beta}=-\frac12(\Box h_{\alpha\beta}
-\partial_\beta \Gamma_\alpha-\partial_\alpha \Gamma_\beta).
\end{equation}
Unlike in Section~\ref{sec:harmvsquo}, 
here $\Box$ is the d'Alembertian associated to the Minkowski metric,
$T_{\alpha\beta}$ is some approximation to the
energy-momentum tensor,
$\Gamma_i$ are the quantities defined in (\ref{eq:3.4}),
and we get from Eq.~(\ref{eq:1.10}):
\begin{equation}
\label{eq:3.15}
\Gamma_0\equiv
-\partial_0h_{00}+\partial_jh_0^j-\frac12\partial_0(-h_{00}+\hat h).
\end{equation}

Assuming that the congruence $u^\alpha$ is harmonic,
that harmonic coordinates are used, $\Gamma_i=0$, and that a foliation with
$\Gamma_0=0$ is selected, then the familiar equations
\begin{equation}
\label{eq:3.16}
R_{\alpha\beta}=-\frac12\Box h_{\alpha\beta}=U_{\alpha\beta}
\end{equation} 
are obtained. These show explicitly the hyperbolic type of Einstein's
linearized field equations when space-time harmonic coordinates can
be used. 

\subsection{Quo-harmonic congruences}\label{sec:linquoharm}

Let us consider now the quantities defined in~(\ref{eq:1.14}):
\begin{equation}
\label{eq:3.9}
\hat\Gamma_i=
\partial_jh^j_i-\frac12\partial_i\hat h.
\end{equation}
Under an adapted coordinate transformation these quantities and their
de\-rivatives with respect to time transform
as 
\begin{equation}
\label{eq:3.10}
\hat\Gamma_{i^\prime}=\Gamma_i+\triangle\zeta_i, \quad 
\partial_0\Gamma_{i^\prime}=\partial_0\Gamma_i.
\end{equation} 
Since at this approximation a congruence is quo-harmonic {\it iff} a system
of coordinates exist such that $\hat\Gamma_{i^\prime}=0$, the condition
\begin{equation}
\label{eq:3.11}
\partial_0\hat\Gamma_j=\partial_i\Sigma^i_j-\frac12\partial_j\Sigma
=0
\end{equation}
characterizes the quo-harmonic congruences to this approximation. 
This is an invariant condition that guarantees the existence
of a solution of the equation
\begin{equation}
\label{eq:3.12}
\triangle\zeta_i=-\hat\Gamma_i
\end{equation}
with $\zeta_i$  independent of $x^0$.

From (\ref{eq:3.7}) and (\ref{eq:3.11}) it follows that a congruence can be
harmonic and quo-harmonic at the same time {\it iff} $\Lambda_i$ is
independent of time. Notice however that this does not mean that in
this case harmonic and quo-harmonic coordinates are the same.
 
Let us assume now that the congruence $u^\alpha$ is  quo-harmonic,
that quo-harmonic coordinates are used, $\hat\Gamma_i=0$,
and that a foliation is selected ---as it can always be done by using (\ref{eq:3.3a})---
such that $\hat\Gamma_0=0$, where
\begin{equation}
\label{eq:3.18}
\hat\Gamma_0 \equiv\partial_ih^i_0.
\end{equation}
The later hypothesis is equivalent to assume that $x^0$ also satisfies the quo-harmonicity
condition~(\ref{eq:2.60}).
Under these assumptions, Einstein's field equations become
\begin{eqnarray}
\label{eq:3.17}
R_{00}&=&-\frac12\left(\triangle h_{00}+\partial_{00}\hat h\right)=U_{00}, \\
\label{eq:3.17b}
R_{0i}&=&-\frac12\left(\triangle h_{0i}+\frac12\partial_{0i}\hat h\right)=U_{0i},
\\
\label{eq:3.17c}
R_{ij}&=&-\frac12\left(\Box h_{ij}+\partial_0(\partial_ih_{0j}+
                       \partial_jh_{0i})-\partial_{ij}h_{00}\right)=U_{ij},
\end{eqnarray}
where $\triangle$ is the Laplacian constructed with the 3-dimensional Euclidean $\delta_{ij}$ metric.
Introducing the traceless tensor
\begin{equation}
\label{eq:3.19}
k_{ij}\equiv h_{ij}-\frac13\hat h\,\delta_{ij},
\end{equation}
the  last group of equations splits in two groups: the scalar equation
\begin{equation}
\label{eq:3.20}
\bar R=-\frac12(\Box\hat h-\triangle h_{00})=\bar U, \quad
\bar R\equiv\delta^{ij}R_{ij}, \quad \bar U \equiv \delta^{ij}U_{ij},
\end{equation}
and the tensor equation
\begin{equation}
\label{eq:3.21}
R_{ij}-\frac13\bar R\delta_{ij}= U_{ij}-\frac13\bar U\delta_{ij}.
\end{equation}
Taking into account Eqs.~(\ref{eq:3.17})--(\ref{eq:3.17c}), 
the scalar Eq.~(\ref{eq:3.20}) reduces to
\begin{equation}
\label{eq:3.22}
-\frac12\triangle\hat h=U_{00}+\bar U
\end{equation}
and the tensor Eq.~(\ref{eq:3.21}) becomes
\begin{equation}
\label{eq:3.23}
-\frac12\left(
\Box k_{ij}+\partial_0(\partial_ih_{0j}+\partial_jh_{0i})-\partial_{ij}h_{00}\right)
-\frac16\triangle h_{00}\delta_{ij}=U_{ij}-\frac13\bar U\delta_{ij}.
\end{equation}

Assuming $T_{\alpha\beta}(x^\rho)$ known everywhere and for all times, 
and satisfying the conservation equations
\begin{equation}
\label{eq:3.24}
\partial_\alpha T^\alpha_\beta=0,
\end{equation}
$\hat h$ can now be obtained integrating the elliptic equation
(\ref{eq:3.22}). After that $h_{00}$ can be obtained integrating the elliptic
equation (\ref{eq:3.17}). And then $h_{0i}$ can be obtained integrating the
elliptic equation (\ref{eq:3.17b}). Finally the single part of the metric
that necessitates to integrate hyperbolic equations is  the traceless piece $k_{ij}$
appearing in  
Eqs~(\ref{eq:3.23}). We can
say then that neither $h_{00}$, nor $h_{0i}$, nor $\hat h$ are
propagating quantities in quo-harmonic coordinates. Only the
wave-part of the metric may propagate. When solving the aforementioned equations
one have to make sure that conditions $\hat\Gamma_0=0$ and $\hat\Gamma_i=0$ are
satisfied, but they are compatible conditions because of conservation law~(\ref{eq:3.24})
and the fact that from the field equations (\ref{eq:3.17})--(\ref{eq:3.17c})
one gets
 \begin{equation}
-\frac12 \triangle\hat\Gamma_0=\partial_\alpha T^\alpha_0 
 \end{equation}
and
 \begin{equation}
-\frac12 \Box\hat\Gamma_i=\partial_\alpha T^\alpha_i+\frac12\partial_{0i}\hat\Gamma_0.
 \end{equation}

\subsection{Stationary solutions}\label{sec:multipoles}

If the frame of reference  in which Eqs.~(\ref{eq:3.16}) or 
Eqs.~(\ref{eq:3.17})--(\ref{eq:3.17c}) have been written is that corresponding to
a Killing congruence implying that $h_{\alpha\beta}$ are time
independent then the preceding equations can be split in two groups. The first group
is common to both harmonic and quo-harmonic coordinates:
\begin{eqnarray}\label{eq:split1}
-\frac12\Delta h_{00}&=&U_{00},\\
-\frac12\Delta h_{0i}&=&U_{0i},\qquad\partial_i h^i_0=0.
\end{eqnarray}
The second group is
\begin{equation}\label{eq:spli2}
-\frac12 \Delta h_{ij}=U_{ij},\qquad
\partial_j h^j_i-\frac12\partial_i\left(-h_{00}+\hat h\right)=0,
\end{equation}
when using harmonic coordinates and 
\begin{equation}\label{eq:spli3}
-\frac12 \left(\Delta h_{ij}-\partial_{ij}h_{00}\right)=U_{ij},\qquad
\partial_j h^j_i-\frac12\partial_i\hat h=0,
\end{equation}
when using quo-harmonic coordinates. The solutions $h_{ij}=\bar h_{ij}$ of 
(\ref{eq:spli2}) and $h_{ij}=\tilde h_{ij}$ of (\ref{eq:spli3})
are related as follows:
\begin{equation}\label{eq:relbarhat}
\tilde h_{ij}=\bar h_{ij}+\partial_i\zeta_j+\partial_j\zeta_i,
\end{equation}
$\zeta_j$ being a solution of the equation
\begin{equation}
\Delta\zeta_j=\frac12\partial_j h_{00}.
\end{equation}

If $U_{\alpha\beta}$ is known, smooth and decreasing fast enough at infinity,
or compact with discontinuities of the first kind
across the border of the support, then the solution to the preceding field equations are:
\begin{equation}
h_{00}=\frac{1}{2\pi}\int{\frac{U_{00}}{R}}\,dV,
\qquad
h_{0i}=\frac{1}{2\pi}\int{\frac{U_{0i}}{R}}\,dV,
\end{equation}
and 
\begin{equation}
\bar h_{ij}=\frac{1}{2\pi}\int{\frac{U_{ij}}{R}}\,dV,
\qquad
\tilde h_{ij}=\frac{1}{4\pi}\int{\frac{2U_{ij}-\partial_{ij}h_{00}}{R}}\,dV,
\end{equation}
As it is well known the solutions thus obtained are at least of class $C^1$. 
This is one of the reasons, among many others, including the analogy with
electromagnetism, to require in general that the potential $g_{\alpha\beta}$
be of class $C^1$ across the surface of discontinuities
of the first kind in the theory of relativity. Surprisingly this requirement
is not respected by many authors because of what appears
to be a confusion between two connected but different concepts:
that of metrics that can be matched and that of metrics that have been matched.

The multipole structure of the potentials $h_{00}$
and $h_{0i}$ can be determined as it is standard in electromagnetic theory.
The multipole structures of $\bar h_{ij}$ and $\tilde h_{ij}$ are in general different
but related by Eq.~(\ref{eq:relbarhat}). Notice also that the multipole structure of 
a particular solution is independent of any system of adapted and admissible coordinates,
harmonic, quo-harmonic or else, if instead of the potentials we consider
the invariant quantities (\ref{eq:inv1})--(\ref{eq:inv4}). 
Furthermore, as a consequence of the field equations (\ref{eq:spli2})--(\ref{eq:spli3}) 
for stationary metrics in vacuum, (\ref{eq:inv4}) is equivalent to
\begin{equation}
\hat R_{ij}=-\frac12\partial_{ij}h_{00}
\end{equation}
both in harmonic and in quo-harmonic coordinates, so that  the invariant
multipole structure in these two coordinate classes for this kind of metrics
is equal and determined by the common values of $h_{00}$ and $h_{0i}$.
In the last section of this paper we shall come back
to the linear approximation in a different context connected with the problem
of defining the multipole structure of a vacuum
solution of Einstein's equations when the source is not known.

\section{Schwarzschild metric}\label{sec:Schw}

To write the Schwarzschild solution we are going to use as basis the 
following forms 
\begin{equation}\label{eq:forms1}
\omega^0=dt, 
\quad\omega^1=dR,
\quad\omega^2=R\,d\Theta,
\quad\omega^3=R\sin\Theta\, d\phi
\end{equation}
 when we use harmonic coordinates and
\begin{equation}\label{eq:forms2}
\omega^0=dt, 
\quad\omega^1= dr,
\quad\omega^2= r\,d\theta,
\quad\omega^3=r\sin\theta\, d\varphi 
\end{equation}
with quo-harmonic coordinates, so that the metric coefficients  $g_{\mu\nu}$
are defined as follows:
\begin{equation}\label{eq:forms3}
ds^2=g_{\mu\nu}\omega^\mu\omega^\nu. 
\end{equation}
These functions reduce to the Minkowskian values 
when $R$ or $r$ go to infinity, 
\begin{equation}\label{eq:forms4}
\lim_{R\mathrm{\ or\ }r\to\infty}g_{\mu\nu}=\eta_{\mu\nu},
\end{equation}
and they will be written as 
\begin{equation}\label{eq:forms5}
g_{\mu\nu}=\eta_{\mu\nu}+h_{\mu\nu},
\end{equation}
where the $h_{\mu\nu}$ are
functions of $R$  or $r$.

\subsection{Schwarzschild metric in harmonic coordinates}\label{sec:Schwharm}

We can obtain a system of harmonic coordinates preserving the
spherical symmetry if we perform a coordinate 
transformation in the form
$(\bar{r}, \bar\theta, \bar\varphi)\rightarrow (R, \Theta,
\phi)$, where $R=R(\bar{r}), \Theta=\bar\theta, \phi=\bar\varphi $.
In fact, the Cartesian coordinates associated to
the set of spherical coordinates $(R,\Theta,\phi)$ are harmonic 
and $\lim_{\bar{r}\rightarrow\infty}R=\bar{r}$ is satisfied if one chooses~\cite{Liu}
\begin{equation}\label{eq:rharschw}
R(\bar{r})=\bar{r}-M+C\left[\frac{\bar{r}-M}{2}
\ln\left(1-\frac{2M}{\bar{r}}\right)+M\right],
\end{equation}
where $C$ is an integration constant, which will be discussed below
in Section~\ref{sec:matching}.

We can invert the coordinate transformation in series of $M/R$ and 
the functions 
$h_{\mu\nu}$ 
of the non-null metric components 
written in harmonic coordinates to 
order five are 
\begin{eqnarray}
h_{00} &= & 2\frac{M}{R}
-2\frac{M^2}{R^2}+2\frac{M^3}{R^3}
-\frac{6+2C}{3}\frac{M^4}{R^4}
+\frac{6+4C}{3}\frac{M^5}{R^5},\nonumber\\[1em]
h_{11} &=&2 \frac{M}{R}+
2\frac{M^2}{R^2}
+\frac{6-4C}{3}\frac{M^3}{R^3}
+\frac{6-10C}{3}\frac{M^4}{R^4}+
\frac{10-28C}{5}\frac{M^5}{R^5},\nonumber\\[1em]
h_{22}&=&h_{33}=2\frac{M}{R}
+\frac{M^2}{R^2}
+\frac23C\frac{M^3}{R^3}
+\frac23C\frac{M^4}{R^4}
+\frac25C\frac{M^5}{R^5}.\label{eq:Schh}
\end{eqnarray}

\subsection{Schwarzschild metric in quo-harmonic coordinates}\label{sec:Schwquo}

We can also obtain a system of quo-harmonic coordinates preserving the
spherical symmetry if we perform a coordinate 
transformation in the form
$(\bar{r}, \bar\theta, \bar\varphi)\rightarrow (r,\theta,\varphi)$, 
with $r=r(\bar{r}),\theta=\bar\theta,\varphi=\bar\varphi$.
The Cartesian coordinates associated to $(r,\theta,\varphi)$
are quo-harmonic and the condition
$\lim_{\bar{r}\rightarrow\infty}r=\bar{r}$ is satisfied if~\cite{jma,Pierre}
\begin{equation}\label{eq:rquoschw}
r(\bar{r})=\bar{r}\left[A\left(1-\frac{3M}{2\bar{r}}\right)+
(1-A)\sqrt{1-\frac{2M}{\bar{r}}}\left(1-\frac{M}{2\bar{r}}\right)\right],
\end{equation}
where $A$ is an integration constant to be discussed later.
Notice the remarkable fact that for $A=0$ the interval $2M\le\bar r<\infty$
corresponds to $0\le r<\infty$. No such thing is possible in harmonic
coordinates.

If we invert the coordinate transformation in series of $M/r$,  
the functions~$h_{\mu\nu}$  of the
non-null metric components in the quo-harmonic coordinates are
to order five  
\begin{eqnarray}
 h_{00}&=&
2\frac{M}{r}
-3\frac{M^2}{r^2}
+\frac{9}{2}\frac{M^3}{r^3} 
-\frac{29-2A}{4}\frac{M^4}{r^4} 
+\frac{99-18A}{8}\frac{M^5}{r^5},\nonumber\\[1em] 
h_{11}& =& 
2\frac{M}{r}
+\frac{M^2}{r^2}
-\frac{1-2A}{2}\frac{M^3}{r^3}   
+\frac{A}{4}\frac{M^4}{r^4}  
-\frac{5 -6A}{8}\frac{M^5}{r^5},\nonumber\\[1em] 
h_{22} &=&h_{33} =
3\frac{M}{r}
+\frac{9}{4}\frac{M^2}{r^2}
+\frac{1-A}{2}\frac{M^3}{r^3}.\label{eq:Schq}  
\end{eqnarray}

\subsection{Matching with an interior solution}\label{sec:matching}

To analyse the meaning of the constants $C$ and $A$ appearing 
in (\ref{eq:rharschw})--(\ref{eq:Schh}) and (\ref{eq:rquoschw})--(\ref{eq:Schq})
respectively, let us consider the problem of constructing a full model for 
the gravitational field of a spherically symmetric star. We can use the
Schwarzschild metric~(\ref{eq:1.1}) for the exterior field but we also need an 
interior solution and then both metrics have to be matched at the star surface.
It is well known that the continuity across the matching surface of
the first and second fundamental forms is enough to guarantee that the matching
can be done without a surface layer~\cite{Israel}. But from our point
of view the problem is not completely solved with this: to actually
make the matching we also need to found for both
metrics a set of ``admissible coordinates'' in the sense of
Lichnerowicz~\cite{Lichne} 
in which all the
derivatives of the metric coefficients are continuous across the star surface. 
Furthermore, we  want this common set of coordinates to have the same physical
meaning for the interior and exterior solutions: they must be members
of a class of systems of coordinates defined with independence of the particular
problem under study. In particular we are interested here in the cases
in which the common coordinates are harmonic 
(or quo-harmonic) for both the interior and the exterior metric.

In fact, Quan-Hui Liu~\cite{Liu} considered a particular interior with uniform density
and found that
the matching with the
Schwarzschild metric~(\ref{eq:1.1}) in harmonic admissible coordinates is only
possible if the constant $C$ in~(\ref{eq:rharschw}) has a given value depending only on the radius
of the star surface.

A similar result arises in quo-harmonic coordinates~\cite{Pierre}:
an interior solution with constant density can be matched in
quo-harmonic admissible coordinates only if the 
constant $A$ in (\ref{eq:rquoschw}) has a particular value
determined by the interior solution, the matching radius and the 
constant~$M$.

These two particular examples suggest that the integration constants
that appear when solving the differential  conditions of
definition for harmonic or quo-harmonic coordinates will be fixed (hopefully
in a unique way) only when the full problem of finding
a common set of such coordinates for both the internal and external gravitational
fields is addressed.

\section{Curzon metric}\label{sec:Curzon}

We are going to write the Curzon metric in the same 
bases~(\ref{eq:forms1})--(\ref{eq:forms2})
we have used for  the Schwarzschild metric.
The metric coefficients $g_{\mu\nu}$ in Eq.~(\ref{eq:forms3}) satisfy 
the asymptotic property~(\ref{eq:forms4}) and we will use again
notation~(\ref{eq:forms5}), but now the $h_{\mu\nu}$ quantities are
functions of $R$ and $\Theta$  or of $r$ and $\theta$.  
Moreover, they have the following structure: 
\begin{equation}
h_{\mu\mu}=\sum_{l}h^{(2l)}_{\mu\mu}P_{2l},\qquad
h_{12}=\sum_{l}h^{(2l)}_{12}P^1_{2l},
\end{equation}
where $h^{(n)}_{\mu\nu}$ are only functions of  $R$ or $r$ and
$P_n$ and $P^1_{n}$ are  the Legendre polynomials  and
associated functions in $\cos\Theta$ or $\cos\theta$.

\subsection{Curzon metric in harmonic coordinates}\label{sec:curzonharm}

One may obtain a system of harmonic coordinates preserving the
axial symmetry 
and the reflection symmetry with respect to the plane $z=0$ 
by using an appropriate coordinate transformation in the form
$(\tilde{r}, \tilde\theta, \tilde\varphi)
\rightarrow (R, \Theta, \phi)$, 
where 
$R= R(\tilde{r}, \tilde\theta),\Theta= \Theta(\tilde{r},\tilde\theta),
\phi=\tilde\varphi$. By solving the
harmonicity equation that
the Cartesian coordinates associated to
the set of spherical coordinates $(R,\Theta,\phi)$ must satisfy,  
and demanding that $\lim_{\tilde{r}\rightarrow\infty}R=\tilde{r}$, 
 we obtain to order five in the expansion parameter $M/R$:
\begin{eqnarray}
\frac{R}{\tilde{r}}&=& 
1
+\frac{\sin^2\tilde\theta}{2}\frac{M^2}{\tilde{r}^2}
-\frac{2C_1+2C_2-3C_2\sin^2\tilde\theta}{6}\frac{M^3}{\tilde{r}^3}
+\frac{4-5\sin^2\tilde\theta}{24}\sin^2\tilde\theta\frac{M^4}{\tilde{r}^4}
\nonumber \\  & &  
+\frac{1}{240}\Big[240C_4+(8C_1+200C_2-420C_3-840C_4)\sin^2\tilde\theta
\nonumber\\&&  \qquad\quad
-(240C_2-525C_3-600C_4)\sin^4\tilde\theta\Big]\frac{M^5}{\tilde{r}^5},
\end{eqnarray}
\begin{eqnarray}
\frac{\cos\Theta}{\cos\tilde\theta} &=&
1
-\frac12\sin^2\tilde\theta\frac{M^2}{\tilde{r}^2}
-\frac{C_2}2\sin^2\tilde\theta\frac{M^3}{\tilde{r}^3}
-\frac{4-11\sin^2\tilde\theta}{24}\sin^2\tilde\theta\frac{M^4}{\tilde{r}^4}
 \nonumber\\ 
 & &
-\frac{1}{80}\Big[
16C_1+80C_2-140C_3-80C_4
\nonumber\\&& \qquad\quad
- (120C_2-175C_3-200C_4)\sin^2\tilde\theta
\Big]\sin^2\tilde\theta\frac{M^5}{\tilde{r}^5},
\end{eqnarray}
where $C_1,\,C_2,\,C_3,C_4$ are integration constants.

In this system of harmonic coordinates 
the functions $h^{(2l)}_{\mu\nu}$  for the non-null components of the
Curzon metric are, to order five,
\begin{eqnarray}
h^{(0)}_{00} &=&   
2\frac{M}{R} 
-2\frac{M^2}{R^2}
+2\frac{M^3}{R^3}
-\frac{6+2C_1}{3}\frac{M^4}{R^4}
+\frac{6+4C_1}{3}\frac{M^5}{R^5},
\nonumber\\[1em] 
h^{(2)}_{00} &=&
-\frac23\frac{M^3}{R^3}
+\frac{4-2C_2}{3}\frac{M^4}{R^4}
-\frac{44-28C_2}{21}\frac{M^5}{R^5},
\nonumber\\[1em] 
h^{(4)}_{00} &=&
\frac{38}{105}\frac{M^5}{R^5},
\\[1em]
h^{(0)}_{11} &=&   
2\frac{M}{R}
+2\frac{M^2}{R^2}
+\frac{6-4C_1}{3}\frac{M^3}{R^3}
+\frac{6-10C_1}{3}\frac{M^4}{R^4}
+\frac{10-28C_1}{5}\frac{M^5}{R^5},\ \ \ %
\nonumber\\[1em] 
h^{(2)}_{11} &=&
-\frac{2+4C_2}3\frac{M^3}{R^3}
-\frac{4+10C_2}{3}\frac{M^4}{R^4}
\nonumber\\&&
-\frac{220 - 168 C_1 + 420 C_2+420C_3 - 360 C_4}{105}\frac{M^5}{R^5},
\nonumber\\[1em]  
h^{(4)}_{11}&=&
\frac{38+420C_3+480C_4}{105}\frac{M^5}{R^5},
\\[1em]
h^{(2)}_{12} &=& 
-\frac{C_2}{6}\frac{M^3}{R^3}
+\frac{2-3C_2}{9}\frac{M^4}{R^4}
\nonumber\\&&
+\frac{140-168C_1-210C_2+420C_3-360C_4}{315}\frac{M^5}{R^5},
\nonumber\\[1em]
h^{(4)}_{12} &=&
\frac{7C_3+8C_4}{56}\frac{M^5}{R^5},
\\[1em]
h^{(0)}_{22} &=& 
2\frac{M}{R}
+\frac{M^2}{R^2}
+\frac{2C_1+C_2}{3}\frac{M^3}{R^3} 
-\frac{1-6C_1-6C_2}{9}\frac{M^4}{R^4} 
\nonumber\\&&
-\frac{40-96C_1-120C_2+105C_3}{180}\frac{M^5}{R^5},
\nonumber\\[1em] 
h^{(2)}_{22} &=& 
-\frac{2+2C_2}{3}\frac{M^3}{R^3}
-\frac{5+18C_2}{9}\frac{M^4}{R^4}
\nonumber\\&&
+\frac{160-1176C_1-3360C_2+1365C_3-4320C_4}{1260}\frac{M^5}{R^5},
\nonumber \\[1em]
h^{(4)}_{22}&=&
\frac{38+315C_3+360C_4}{105}\frac{M^5}{R^5},
\\[1em]
h^{(0)}_{33} &=&   
2\frac{M}{R}
+\frac{M^2}{R^2}
+\frac{2C_1-C_2}{3}\frac{M^3}{R^3}
+\frac{1+6C_1-6C_2}{9}\frac{M^4}{R^4}
\nonumber\\&&
+\frac{40+48C_1-120C_2+105C_3}{180}\frac{M^5}{R^5},
\nonumber\\[1em]
h^{(2)}_{33} &=&
-\frac23\frac{M^3}{R^3}
-\frac{7+6C_2}{9}\frac{M^4}{R^4}
-\frac{80+168C_1+336C_2-735C_3}{252}\frac{M^5}{R^5},
\nonumber\\[1em]
h^{(4)}_{33}&=& \frac{38}{105} \frac{M^5}{R^5}. 
\end{eqnarray}

\subsection{Curzon metric in quo-harmonic coordinates}\label{sec:curzonquo}

To obtain a system of quo-harmonic coordinates preserving the
axial  and reflection symmetries we will perform
a coordinate transformation in the form $(\tilde{r}, \tilde\theta, \tilde\varphi)
\rightarrow (r, \theta, \varphi)$,
with 
$r= r(\tilde{r}, \tilde\theta),\theta= \theta(\tilde{r},\tilde\theta),
\varphi=\tilde\varphi$. By demanding $\lim_{\tilde{r}\rightarrow\infty}r=\tilde{r}$
and that the Cartesian coordinates associated to the spherical $(r,\theta,\varphi)$
satisfy the quo-harmonicity equation
we obtain to order five in $M/r$
\begin{eqnarray}
\frac{r}{\tilde{r}}&=&
1
-\frac12\frac{M}{\tilde{r}}
+\frac{\sin^2\tilde{\theta}}{2}\frac{M^2}{\tilde{r}^2}
+\frac{1+A_1-2A_2-(3-3A_2)\sin^2\tilde\theta}{4}\frac{M^3}{\tilde{r}^3}
\nonumber\\&&
-\frac{3+3A_1-6 A_2-(13-9A_2)\sin^2\tilde\theta+5\sin^4\tilde\theta}{24}
\frac{M^4}{\tilde{r}^4}
\nonumber\\&&
+\frac{1}{240}\Big[9+9A_1-18A_2+6A_4
\nonumber\\&&\qquad\quad
+(59-6A_1+327A_2-288A_3-21A_4)\sin^2\tilde\theta
\nonumber\\&&\qquad\quad
-(115+360A_2-360A_3-15A_4)\sin^4\tilde\theta
\Big]\frac{M^5}{\tilde{r}^5},\ \ \ %
\end{eqnarray}
\begin{eqnarray}
\frac{\cos\theta}{\cos\tilde\theta} & = & 
1
-\frac{\sin^2\tilde\theta}{2}\frac{M^2}{\tilde{r}^2}
+\frac{2-3A_2}{4}\sin^2\tilde\theta\frac{M^3}{\tilde{r}^3}
-\frac{7-11\sin^2\tilde\theta}{24}\sin^2\tilde\theta\frac{M^4}{\tilde{r}^4}
\nonumber\\&&
-\frac{1}{240}\Big[64-36A_1+387A_2-288A_3-6A_4
\nonumber\\&&\qquad\quad
-(20+540A_2-360A_3-15A_4)\sin^2\tilde\theta
\Big]\sin^2\tilde\theta\frac{M^5}{\tilde{r}^5},
\end{eqnarray}
where $A_1$, $A_2$, $A_3$ and $A_4$ are integration constants.

In this system of quo-harmonic coordinates the functions $h^{(2l)}_{\mu\nu}$  
for the  non-null components of Curzon  metric are written as follows:
\begin{eqnarray}
h^{(0)}_{00} &=&   
2\frac{M}{r} 
-3\frac{M^2}{r^2}
+\frac92\frac{M^3}{r^3}
-\frac{29-2A_1}{4}\frac{M^4}{r^4}
+\frac{99-18A_1}{8}\frac{M^5}{r^5},
\nonumber\\[1em] 
h^{(2)}_{00} &=&
-\frac23\frac{M^3}{r^3}
+\frac{10-3A_2}{3}\frac{M^4}{r^4}
-\frac{431-189A_2}{42}\frac{M^5}{r^5},
\nonumber\\[1em] 
h^{(4)}_{00} &=&
\frac{38}{105}\frac{M^5}{r^5},
\\[1em]
h^{(0)}_{11} &=&   
2\frac{M}{r}
+\frac{M^2}{r^2}
-\frac{1-2A_1}{2}\frac{M^3}{r^3}
+\frac{A_1}{4}\frac{M^4}{r^4}
-\frac{5-6A_1}{8}\frac{M^5}{r^5},\ \ \ %
\nonumber\\[1em] 
h^{(2)}_{11} &=&
\frac{4-6A_2}3\frac{M^3}{r^3}
+\frac{1-3A_2}{6}\frac{M^4}{r^4}
\nonumber\\&&
+\frac{733 - 252 A_1 + 189 A_2-576A_3 +18 A_4}{210}\frac{M^5}{r^5},
\nonumber\\[1em]  
h^{(4)}_{11}&=&
-\frac{114-96A_3-4A_4}{35}\frac{M^5}{r^5},
\\[1em]
h^{(2)}_{12} &=& 
-\frac{A_2}{4}\frac{M^3}{r^3}
+\frac{11-18A_2}{36}\frac{M^4}{r^4}
\nonumber\\&&
-\frac{2002-1008A_1+1701A_2-2304A_3+72A_4}{2520}\frac{M^5}{r^5},
\nonumber\\[1em]
h^{(4)}_{12} &=&
\frac{24A_3+A_4}{280}\frac{M^5}{r^5},
\\[1em]
h^{(0)}_{22} &=& 
3\frac{M}{r}
+\frac94\frac{M^2}{r^2}
+\frac{1-3A_1+3A_2}{6}\frac{M^3}{r^3} 
-\frac{19-27A_2}{36}\frac{M^4}{r^4} 
\nonumber\\&&
+\frac{28-9A_1+18A_2-36A_3}{90}\frac{M^5}{r^5},
\nonumber\\[1em] 
h^{(2)}_{22} &=& 
-\frac{1+3A_2}{3}\frac{M^3}{r^3}
+\frac{13-27A_2}{9}\frac{M^4}{r^4}
\nonumber\\&&
-\frac{94-441A_1+882A_2-468A_3+54A_4}{630}\frac{M^5}{r^5},
\nonumber \\[1em]
h^{(4)}_{22}&=&
-\frac{57-72A_3-3A_4}{35}\frac{M^5}{r^5},
\\[1em]
h^{(0)}_{33} &=&   
3\frac{M}{r}
+\frac94\frac{M^2}{r^2}
+\frac{5-3A_1-3A_2}{6}\frac{M^3}{r^3}
+\frac{19-27A_2}{36}\frac{M^4}{r^4}
\nonumber\\&&
-\frac{28-9A_1+18A_2-36A_3}{90}\frac{M^5}{r^5},
\nonumber\\[1em]
h^{(2)}_{33} &=&
-\frac{M^3}{r^3}
+\frac{7-27A_2}{18}\frac{M^4}{r^4}
-\frac{214-63A_1+126A_2-252A_3}{126}\frac{M^5}{r^5},
\nonumber\\[1em]
h^{(4)}_{33}&=& 
\frac{19}{35} \frac{M^5}{r^5}. 
\end{eqnarray}

Since now we have two sets of coordinates that can be considered
common to the Schwarzschild and Curzon metrics, we can compare the
two metrics. From the results in Section~\ref{sec:Schw} we see that
both metrics coincide if terms of order $M^3/R^3$ (or $M^3/r^3$)
and higher are neglected.
This means that in most
practical cases (such as in relativistic celestial mechanics) these
two metrics are indistinguishable if used in harmonic
or quo-harmonic coordinates. 
Notice that if one naively identifies
the coordinates used in (\ref{eq:1.1}) and (\ref{eq:1.2})
the difference between the Schwarzschild and Curzon metrics 
starts with terms proportional to $M/\bar r=M/\tilde r$.
These remarks have no special utility except that
they demystify the apparent weirdness of the Curzon solution.

\section{Kerr metric}\label{sec:Kerr}

To write the Kerr metric  we will use again~(\ref{eq:forms1})--(\ref{eq:forms3}) and
the metric coefficients $g_{\mu\nu}$ will satisfy~(\ref{eq:forms4}).
The $h_{\mu\nu}$ quantities of~(\ref{eq:forms5})  are
functions of $R$ and $\Theta$  or of $r$ and $\theta$ and 
may be written as
\begin{equation}
h_{\mu\mu}=\sum_{l}h^{(2l)}_{\mu\mu}P_{2l},\quad
h_{12}=\sum_{l}h^{(2l)}_{12}P^1_{2l},\quad
h_{03}=\sum_{l}h^{(2l+1)}_{03}P^1_{2l+1},
\end{equation}
where $h^{(n)}_{\mu\nu}$ are only functions of  $R$ or $r$ and
$P_n$ and $P^1_{n}$ are the Legendre polynomials and 
associated functions in $\cos\Theta$ or $\cos\theta$.

We will use the dimensionless quantity
\begin{equation}
\alpha\equiv\frac{a}{M}
\end{equation}
to make easier the comparison with the results for the Schwarzschild metric,
which is recovered in the limit $\alpha=0$.

\subsection{Kerr metric in harmonic coordinates}\label{sec:kerrharm}

To preserve the axial symmetry and the reflection symmetry 
with respect to the plane $z=0$ we perform a coordinate transformation 
$(\check{r}, \check\theta, \check\varphi)
\rightarrow (R, \Theta, \phi)$, 
with
$R= R(\check{r}, \check\theta),\Theta= \Theta(\check{r},\check\theta),
\phi=\check\varphi$. If  $\lim_{\check{r}\rightarrow\infty}R=\check{r}$
and  the Cartesian coordinates associated to
the spherical $(R,\Theta,\phi)$ are harmonic, 
the transformation is given to order five in $1/R$ (now we have
two dimensionless expansion parameters: $M/R$ and $a/R$) by 
\begin{eqnarray}
\frac{R}{\check{r}} &=& 
1
-\frac{M}{\check{r}}
+\frac{\alpha^2}{2}\sin^2\check\theta\frac{M^2}{\check{r}^2}
-\frac{2D_1+\alpha^2D_2\left(2-3\sin^2\check\theta\right)}{6}\frac{M^3}{\check{r}^3}
\nonumber\\&&
-\frac{16D_1+ 4\alpha^2\left[4D_2+(3-6D_2)\sin^2\check\theta\right]+
3\alpha^4\sin^4\check\theta}{24}\frac{M^4}{\check{r}^4}
\nonumber\\&&
-\frac{1}{120}\Big[
144D_1+30\alpha^2\left(2D_2+D_3-16D_4\right)
\nonumber\\&&\qquad\quad
+5\alpha^2\left(
156-8D_1-18D_2-9D_3+480D_4-20\alpha^2D_2
\right)\sin^2\check\theta
\nonumber\\&&\qquad\quad
-60\alpha^2\left(
13+35D_4-2\alpha^2D_2
\right)\sin^4\check\theta
\Big]\frac{M^5}{\check{r}^5},
\end{eqnarray}
\begin{eqnarray}
\frac{\cos\Theta}{\cos\check\theta} &=& 
1
-\frac{\alpha^2}{2}\sin^2\check\theta\frac{M^2}{\check{r}^2}
-\alpha^2\frac{1+D_2}{2}\sin^2\check\theta\frac{M^3}{\check{r}^3}
\nonumber\\&&
-\alpha^2\frac{12D_2-3\alpha^2\sin^2\check\theta}{8}\sin^2\check\theta\frac{M^4}{\check{r}^4}
\nonumber\\&&
+\frac{\alpha^2}{8}\Big[
52-32D_2+2D_3+80D_4-4\alpha^2D_2
\nonumber\\&&\qquad\quad
-\left(
52+140D_4-\alpha^2(5+12D_2)
\right)\sin^2\check\theta
\Big]\sin^2\check\theta\frac{M^5}{\check{r}^5},\ \ \ \ \ \ %
\end{eqnarray}
where $D_1,D_2,D_3,D_4$ are integration constants.

In this system of harmonic coordinates 
the functions $h^{(l)}_{\mu\nu}$  for the
non-null components of the Kerr   metric are written as follows to order five:
\begin{eqnarray}
h^{(0)}_{00} &=&   
2\frac{M}{R} 
-2\frac{M^2}{R^2}
+2\frac{M^3}{R^3}
-\frac{6+2D_1}{3}\frac{M^4}{R^4}
+\frac{6+4D_1-2\alpha^2}{3}\frac{M^5}{R^5},
\nonumber\\[1em] 
h^{(2)}_{00} &=&
-2\alpha^2\frac{M^3}{R^3}
+\alpha^2\frac{18-2D_2}{3}\frac{M^4}{R^4}
-\alpha^2\frac{34-4D_2}{3}\frac{M^5}{R^5},
\nonumber\\[1em] 
h^{(4)}_{00} &=&
2\alpha^4\frac{M^5}{R^5},
\\[1em]
h^{(0)}_{11} &=&   
2\frac{M}{R}
+2\frac{M^2}{R^2}
+\frac{6-4D_1-4\alpha^2}{3}\frac{M^3}{R^3}
+\frac{6-10D_1-10\alpha^2}{3}\frac{M^4}{R^4}
\nonumber\\&&
+\frac{30-84D_1-94\alpha^2}{15}\frac{M^5}{R^5},\ \ \ %
\nonumber\\[1em] 
h^{(2)}_{11} &=&
-\alpha^2\frac{2+4D_2}3\frac{M^3}{R^3}
+\alpha^2\frac{4-10D_2}{3}\frac{M^4}{R^4}
-\alpha^2\frac{202 +42D_3 - 24\alpha^2}{21}\frac{M^5}{R^5},
\nonumber\\[1em]  
h^{(4)}_{11}&=&
\alpha^2\frac{416+1120D_4+30\alpha^2}{35}\frac{M^5}{R^5},
\\[1em]
h^{(2)}_{12} &=& 
\alpha^2\frac{1-D_2}{6}\frac{M^3}{R^3}
+\alpha^2\frac{2-D_2}{3}\frac{M^4}{R^4}
+\alpha^2\frac{100-42D_2+14D_3-8\alpha^2}{21}\frac{M^5}{R^5},
\nonumber\\[1em]
h^{(4)}_{12} &=&
\alpha^2\frac{52+140D_4-5\alpha^2}{140}\frac{M^5}{R^5},
\\[1em]
h^{(0)}_{22} &=& 
2\frac{M}{R}
+\frac{M^2}{R^2}
+\frac{2D_1+\alpha^2(1+D_2)}{3}\frac{M^3}{R^3} 
+\frac{2D_1+\alpha^2(1+2D_2)}{3}\frac{M^4}{R^4} 
\nonumber\\&&
+\frac{12D_1-\alpha^2(56-30D_2+5D_3+60D_4)+5\alpha^4}{30}\frac{M^5}{R^5},
\nonumber\\[1em] 
h^{(2)}_{22} &=& 
-\alpha^2\frac{4+2D_2}{3}\frac{M^3}{R^3}
+\alpha^2\frac{2-6D_2}{3}\frac{M^4}{R^4}
\nonumber\\&&
+\alpha^2\frac{208-210D_2+49D_3-420D_4-13\alpha^2}{42}\frac{M^5}{R^5},
\nonumber \\[1em]
h^{(4)}_{22}&=&
\alpha^2\frac{312+840D_4+40\alpha^2}{35}\frac{M^5}{R^5},
\\[1em]
h^{(0)}_{33} &=&   
2\frac{M}{R}
+\frac{M^2}{R^2}
+\frac{2D_1+\alpha^2(3-D_2)}{3}\frac{M^3}{R^3}
+\frac{2D_1-\alpha^2(1+2D_2)}{3}\frac{M^4}{R^4}
\nonumber\\&&
+\frac{12D_1+5\alpha^2(24-6D_2+D_3+12D_4)-5\alpha^4}{30}\frac{M^5}{R^5},
\nonumber\\[1em]
h^{(2)}_{33} &=&
-2\alpha^2\frac{M^3}{R^3}
+\alpha^2\frac{4-2D_2}{3}\frac{M^4}{R^4}
\nonumber\\&&
+\alpha^2\frac{48-18D_2+5D_3+60D_4-5\alpha^2}{6}\frac{M^5}{R^5},
\nonumber\\[1em]
h^{(4)}_{33}&=& 
2\alpha^4 \frac{M^5}{R^5}, \\[1em]
h^{(1)}_{03} &=&
2\alpha\frac{M^2}{R^2}
-2\alpha\frac{M^3}{R^3}
+2\alpha\frac{M^4}{R^4}
-\alpha\frac{30+10D_1+\alpha^2(6+4D_2)}{15}\frac{M^5}{R^5},
\nonumber\\[1em]
h^{(3)}_{03}&=& 
-\frac23\alpha^3 \frac{M^4}{R^4}
+\alpha^3\frac{26-6D_2}{15}\frac{M^5}{R^5}. 
\end{eqnarray}

\subsection{Kerr metric in quo-harmonic coordinates}\label{sec:kerrquo}

We can also obtain a system of quo-harmonic coordinates preserving the
axial and specular symmetries by using a change of coordinate in the form
$(\check{r}, \check\theta, \check\varphi)
\rightarrow (r, \theta, \varphi)$, 
with 
$r= r(\check{r}, \check\theta),\theta= \theta(\check{r},\check\theta),
\varphi=\check\varphi$. The Cartesian coordinates associated to
$(r,\theta,\varphi)$ will be quo-harmonic  
and $\lim_{\check{r}\rightarrow\infty}r=\check{r}$ will be satisfied if,
up to order $1/r^5$, we choose
\begin{eqnarray}
\frac{r}{\check{r}} &=& 
1
-\frac32\frac{M}{\check{r}}
+\frac{\alpha^2}{2}\sin^2\check\theta\frac{M^2}{\check{r}^2}
-\frac{3-3B_1-\alpha^2B_2\left(4-6\sin^2\check\theta\right)}{12}\frac{M^3}{\check{r}^3}
\nonumber\\&&
-\frac{3-3B_1-2\alpha^2\left[2B_2+(1-3B_2)\sin^2\check\theta\right]+
\alpha^4\sin^4\check\theta}{8}\frac{M^4}{\check{r}^4}
\nonumber\\&&
-\frac{6}{35}\alpha^2\left(2-3\sin^2\check\theta\right)\frac{M^5}{\check{r}^5}\ln\frac{M}{\check{r}}
\nonumber\\&&
-\frac{1}{3360}\Big[
1890-1890B_1-12\alpha^2\left(38+175B_2+96B_3+2B_4\right)+840\alpha^4
\nonumber\\&&\qquad\quad
-2\alpha^2\Big(
1044-420B_1-1575B_2-864B_3-60B_4
\nonumber\\&&\qquad\qquad\qquad
+140\alpha^2(9-10B_2)
\Big)\sin^2\check\theta
\nonumber\\&&\qquad\quad
-105\alpha^2\left(
B_4-\alpha^2(15-32B_2)\right)\sin^4\check\theta
\Big]\frac{M^5}{\check{r}^5},
\end{eqnarray}
\begin{eqnarray}
\frac{\cos\theta}{\cos\check\theta} &=& 
1
-\frac{\alpha^2}{2}\sin^2\check\theta\frac{M^2}{\check{r}^2}
-\alpha^2\frac{5-2B_2}{4}\sin^2\check\theta\frac{M^3}{\check{r}^3}
\nonumber\\&&
-\alpha^2\frac{17-12B_2-3\alpha^2\sin^2\check\theta}{8}\sin^2\check\theta\frac{M^4}{\check{r}^4}
\nonumber\\&&
+\frac{12}{35}\alpha^2\sin^2\check\theta\frac{M^5}{\check{r}^5}\ln\frac{M}{\check{r}}
\nonumber\\&&
-\frac{\alpha^2}{1120}\Big[
3694-3920B_2+384B_3-20B_4
-140\alpha^2(3+4B_2)
\nonumber\\&&\qquad\quad
+\left(
35B_4-35\alpha^2(37-48B_2)
\right)\sin^2\check\theta
\Big]\sin^2\check\theta\frac{M^5}{\check{r}^5},\ \ \ \ \ \ %
\end{eqnarray}
where we have again four integration constants: $B_1$, $B_2$, $B_3$
and $B_4$.

Notice that, unlike in all the preceding examples, here we do not
have only powers of $1/r$ but also terms of the form $\ln r/r^5$.

Functions $h^{(l)}_{\mu\nu}$  for the non-null components 
of Kerr metric appear as follows in this system of quo-harmonic coordinates:
\begin{eqnarray}
h^{(0)}_{00} &=&   
2\frac{M}{r} 
-3\frac{M^2}{r^2}
+\frac92\frac{M^3}{r^3}
-\frac{29-2B_1}{4}\frac{M^4}{r^4}
+\frac{297-54B_1+8\alpha^2}{24}\frac{M^5}{r^5},
\nonumber\\[1em] 
h^{(2)}_{00} &=&
-2\alpha^2\frac{M^3}{r^3}
+\alpha^2\frac{27+2B_2}{3}\frac{M^4}{r^4}
-\alpha^2\frac{82+9B_2}{3}\frac{M^5}{r^5},
\nonumber\\[1em] 
h^{(4)}_{00} &=&
2\alpha^4\frac{M^5}{r^5},
\\[1em]
h^{(0)}_{11} &=&   
2\frac{M}{r}
+\frac{M^2}{r^2}
-\frac{3-6B_1+8\alpha^2}{6}\frac{M^3}{r^3}
+\frac{3B_1+20\alpha^2}{12}\frac{M^4}{r^4}
\nonumber\\&&
-\frac{25-30B_1+104\alpha^2}{40}\frac{M^5}{r^5},\ \ \ %
\nonumber\\[1em] 
h^{(2)}_{11} &=&
-\alpha^2\frac{2-4B_2}3\frac{M^3}{r^3}
-\alpha^2\frac{2-B_2}{3}\frac{M^4}{r^4}
-\frac{96}{35}\alpha^2\frac{M^5}{r^5}\ln\frac{M}{r}+\frac{96}{35}\alpha^2B_3\frac{M^5}{r^5},
\nonumber\\[1em]  
h^{(4)}_{11}&=&
\frac{2}{35}\alpha^2B_4\frac{M^5}{r^5},
\\[1em]
h^{(2)}_{12} &=& 
-\alpha^2\frac{7-2B_2}{12}\frac{M^3}{r^3}
+\alpha^2\frac{5+4B_2}{12}\frac{M^4}{r^4}
+\alpha^2\frac{32}{35}\frac{M^5}{r^5}\ln\frac{M}{r}
\nonumber\\&&
-\alpha^2\frac{61-210B_2+768B_3}{840}\frac{M^5}{r^5},
\nonumber\\[1em]
h^{(4)}_{12} &=&
\alpha^2\frac{B_4+315\alpha^2}{560}\frac{M^5}{r^5},
\\[1em]
h^{(0)}_{22} &=& 
3\frac{M}{r}
+\frac94\frac{M^2}{r^2}
+\frac{3-3B_1+\alpha^2(5-2B_2)}{6}\frac{M^3}{r^3} 
-\alpha^2\frac{1+3B_2}{6}\frac{M^4}{r^4} 
\nonumber\\&&
-\frac{8}{35}\alpha^2\frac{M^5}{r^5}\ln\frac{M}{r}
+\alpha^2\frac{48-70B_2+192B_3-3B_4-105\alpha^2}{840}\frac{M^5}{r^5},
\nonumber\\[1em] 
h^{(2)}_{22} &=& 
-\alpha^2\frac{10-2B_2}{3}\frac{M^3}{r^3}
-\alpha^2\frac{1-6B_2}{3}\frac{M^4}{r^4}
\nonumber\\&&
+\frac{8}{5}\alpha^2\frac{M^5}{r^5}\ln\frac{M}{r}
+\alpha^2\frac{168+490B_2-1344B_3-15B_4-525\alpha^2}{840}\frac{M^5}{r^5},
\nonumber \\[1em]
h^{(4)}_{22}&=&
\alpha^2\frac{3B_4+315\alpha^2}{70}\frac{M^5}{r^5},
\\[1em]
h^{(0)}_{33} &=&   
3\frac{M}{r}
+\frac94\frac{M^2}{r^2}
+\frac{3-3B_1+\alpha^2(3+2B_2)}{6}\frac{M^3}{r^3}
-\alpha^2\frac{5-B_2}{2}\frac{M^4}{r^4}
\nonumber\\&&
+\frac{8}{35}\alpha^2\frac{M^5}{r^5}\ln\frac{M}{r}
+\alpha^2\frac{2864+70B_2-192B_3+3B_4+105\alpha^2}{840}\frac{M^5}{r^5},
\nonumber\\[1em]
h^{(2)}_{33} &=&
-3\alpha^2\frac{M^3}{r^3}
+\alpha^2(2+B_2)\frac{M^4}{r^4}
+\frac{8}{7}\alpha^2\frac{M^5}{r^5}\ln\frac{M}{r}
\nonumber\\&&
-\alpha^2\frac{2648-350B_2+960B_3-15B_4-525\alpha^2}{840}\frac{M^5}{r^5},
\nonumber\\[1em]
h^{(4)}_{33}&=& 
3\alpha^4 \frac{M^5}{r^5}, \\[1em]
h^{(1)}_{03} &=&
2\alpha\frac{M^2}{r^2}
-3\alpha\frac{M^3}{r^3}
+\frac92\alpha\frac{M^4}{r^4}
-\alpha\frac{435-30B_1+4\alpha^2(15-4B_2)}{60}\frac{M^5}{r^5},
\nonumber\\[1em]
h^{(3)}_{03}&=& 
-\frac23\alpha^3 \frac{M^4}{r^4}
+\alpha^3\frac{35+6B_2}{15}\frac{M^5}{r^5}. 
\end{eqnarray}

We may now compare the Schwarzschild, Curzon and Kerr metrics
in a common (harmonic or quo-harmonic) set of coordinates defined
with independence of the models being analysed. If we neglect
again  terms of order $1/R^3$ (or $1/r^3$) and higher, we see
that the only difference between the two static cases and the 
Kerr metric is the term $h^{(1)}_{03}=2Ma/R^2$ 
(or $h^{(1)}_{03}=2Ma/r^2$).

\section{Final comments}\label{sec:final}

The approximation scheme in Sections.~\ref{sec:Schw}, \ref{sec:Curzon}
and~\ref{sec:Kerr} is based on expanding 
an exact solution of the complete 
Einstein's equations about the infinity point of an appropriate radial coordinate. 
In the so-called linear approximation discussed in Section~\ref{sec:linear}
the exact solutions of the approximated (linearized) field equations were used.
Nevertheless, by carefully selecting some terms from the expansions 
in Sections.~\ref{sec:Schw}, \ref{sec:Curzon} and~\ref{sec:Kerr} one may recover
exact solutions of the linearized theory. 

For instance,
from the expressions in Section~\ref{sec:Schwharm} for the Schwarzschild
metric in harmonic coordinates, one may easily find the following solution
of the linear approximation:
\begin{eqnarray}
h_{00}&=&\frac{2M}R,\\
h_{0i}&=&0.
\end{eqnarray}
The same happens (after replacing $R$ with $r$) 
in the expressions in Section~\ref{sec:Schwquo} for the Schwarzschild
metric in quo-harmonic coordinates. Notice that we are here writing only the 
metric components that give the common invariant multipole structure,
as discussed in Section~\ref{sec:multipoles}.

In the case of the expressions in Section~\ref{sec:curzonharm} for the Curzon
metric in harmonic coordinates, one may easily find the following solution
of the linear approximation with more complex angular structure:
\begin{eqnarray}
h_{00}&=&\frac{2M}{R}
-\frac{2}{3}\frac{M^3}{R^3}P_2\left(\cos\Theta\right)
+\frac{38}{105}\frac{M^5}{R^5}P_4\left(\cos\Theta\right),\\
h_{0i}&=&0.
\end{eqnarray}
Substituting $r$ for $R$ and $\theta$ for $\Theta$ in this result we get the solution 
of the linearized field equations
contained in the approximated solution of Section~\ref{sec:curzonquo}
for quo-harmonic coordinates.

In the case of the Kerr metric, the solutions of the linearized
theory  we can extract from  Section~\ref{sec:Kerr} are
\begin{eqnarray}
h_{00}&=&
\frac{2M}{R}
-2\frac{Ma^2}{R^3}P_2\left(\cos\Theta\right)
+2\frac{Ma^4}{R^5}P_4\left(\cos\Theta\right),\\
h_{01}&=&h_{02}=0,\\
h_{03}&=&4\frac{Ma}{R^2}P^1_1\left(\cos\Theta\right)
-\frac43\frac{Ma^3}{R^4}P^1_3\left(\cos\Theta\right)
\end{eqnarray}
in harmonic coordinates, as well as in quo-harmonic coordinates if one uses $(r,\theta)$ 
instead of $(R,\Theta)$.

\section*{Acknowledgments}

The work of JMA was supported by the University of the Basque Country
through the research project~UPV172.310-EB150/98 and the General Research
Grant~UPV172.310-G02/99. Ll.~Bel gratefully acknowledges
as visiting professor the hospitality of the UPV/EHU. AM was supported by the
research projects CICYT BF2000-0604 and CIRIT 2000SGR-00023. JM and ER were
supported by the research project BFM2000 - 1322
(Ministerio de Ciencia y Tecnologia, Spain)


\begin{thebibliography}{99}

\bibitem{Tolo} Ll.\ Bel and B.\ Coll, Gen.\ Rel.\ Grav.\ \textbf{25}, 613 (1993).

\bibitem{Salas} Ll.\ Bel in \textit{Relativity in General},
J.\ D\'\i az and M.\ Lorente, Eds. Editions Fronti\`eres, p.\ 47 (1994).

\bibitem{Llosa1} Ll.\ Bel and J.\ Llosa,  Gen.\ Rel.\ and Grav.\ \textbf{27}, 1089 (1995).

\bibitem{Llosa2} Ll.\ Bel and J.\ Llosa, Class.\ and Quant.\ Grav.\ \textbf{12}, 1949 (1995).

\bibitem{bers} L.\ Bers, F.\ John and M.\ Schechter, \textit{Partial Differential Equations},
Wiley, New York (1964).

\bibitem{deturk} D.\ M.\ Deturk and L.\ Kazdan, Ann.\ Sci.\ Ecole Norm.\ Sup.\ 4e s\'erie,
\textbf{14}, 249 (1981).

\bibitem{zelmanov} A.\ L.\ Zel'manov, Sov.\ Phys.\ Dokl.\ \textbf{1}, 227 (1956).

\bibitem{cattaneo} C.\ Cattaneo, Ann.\ di Mat.\ pura ed appli., S.\ IV, T.\ XLVIII. 361 (1959).

\bibitem{Liu}  Q.-H.\ Liu, J.\ Math.\ Phys.\ \textbf{39}, 6086 (1998).

\bibitem{jma} J.\ M.\ Aguirregabiria, unpublished.

\bibitem{Pierre} P.\ Teyssandier, private communication.

\bibitem{Israel} W.\ Israel, Il Nuovo Cimento, \textbf{XLIV B}, 1 (1966).

\bibitem{Lichne} A.\ Lichnerowicz, \textit{Th\'eories relativistes
de la Gravitation et de l' Electromagnetisme}, Masson, Paris (1955). 

\end{thebibliography}
\end{document}